\documentclass{ws-procs9x6}

\begin{document}

\title{Restoration of chiral and $U(1)_A$ symmetries in excited hadrons in the semiclassical regime}

\author{L. Ya. GLOZMAN}

\address{Institute for physics/Theoretical physics, University of Graz,\\
Universit\"atsplatz 5, A-8010 Graz, Austria\\
$^*$E-mail: leonid.glozman@uni-graz.at}

\begin{abstract}
Restoration of chiral and $U(1)_A$ symmetries in excited hadrons is
reviewed. Implications of the OPE as well as of the semiclassical expansion
for this phenomenon are discussed. A solvable model of the 't Hooft type in 3+1 
dimensions is presented, which demonstrates a fast restoration of both chiral and $U(1)_A$
symmetries at larger spins and radial excitations.
\end{abstract}


\bodymatter

\section{Introduction}\label{sec1}
There are some phenomenological evidences that the highly excited hadrons,
both baryons \cite{G1,CG,G2} and mesons \cite{G3,G4} fall into approximate
multiplets of $SU(2)_L \times SU(2)_R$ and $U(1)_A$ groups, for a short overview see 
ref. \cite{G5}.
This is illustrated in Fig. 1, where the excitation spectrum of the nucleon as well as
the excitation spectrum of $\pi$ and $f_0$ (with the $\bar n n = \frac{\bar u u + \bar d d}{\sqrt 2}$
content) mesons  are shown. Starting from the 1.7 GeV
region the nucleon (and delta) spectra show obvious signs of
parity doubling. There are a couple of examples where chiral
partners of highly excited states have not yet been seen. Their
experimental discovery would be an important task. Similarly, in
the chirally restored regime $\pi$ and $\bar n n$ $f_0$ states
must be systematically degenerate. This phenomenon, if experimentally
confirmed by discovery of still missing states, is referred
to as effective chiral symmetry restoration or chiral symmetry
restoration of the second kind. 

\begin{figure}

\begin{center}
\includegraphics*[width=4cm,angle=-90]{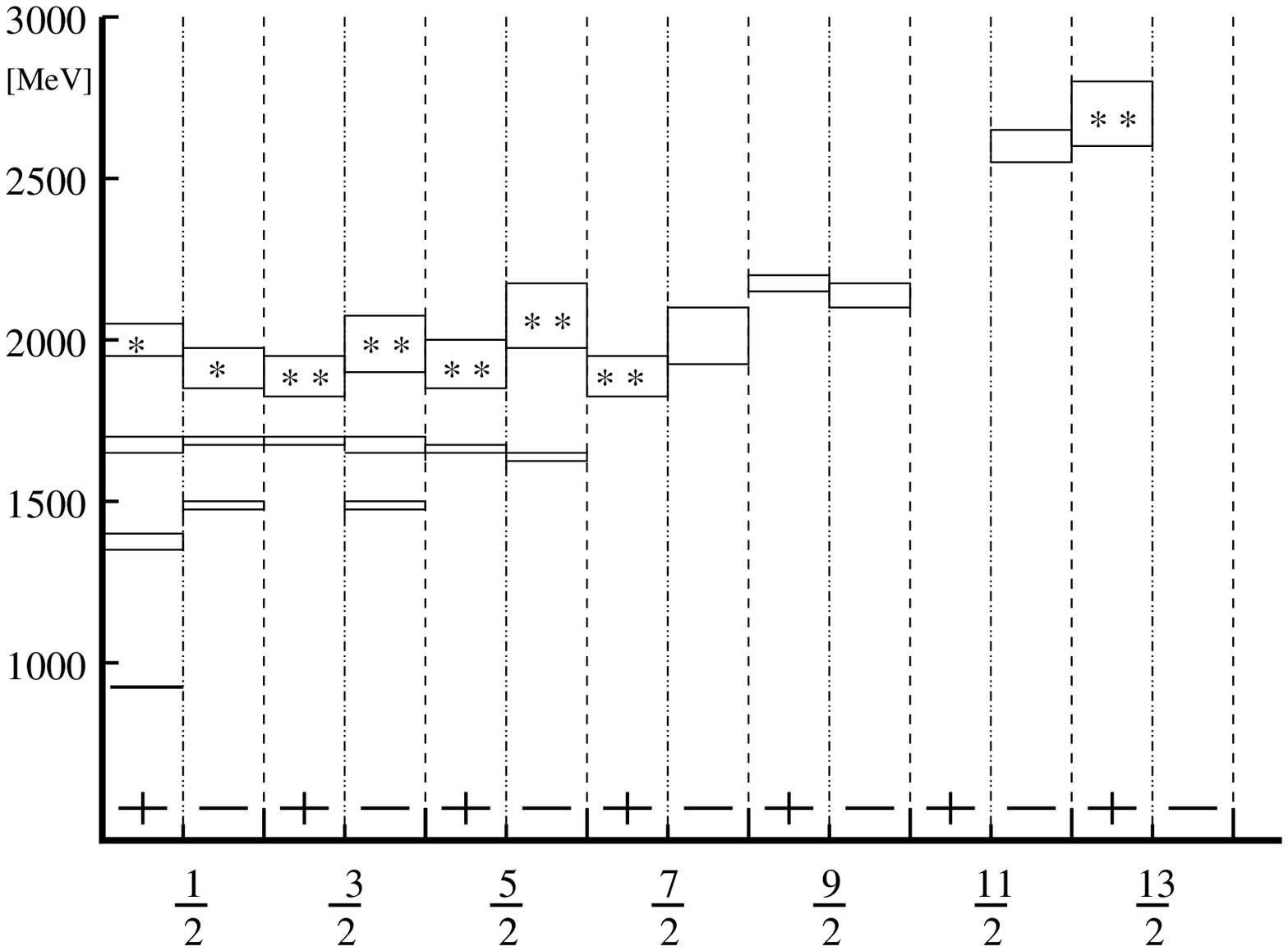}
\includegraphics*[width=4cm,angle=-90]{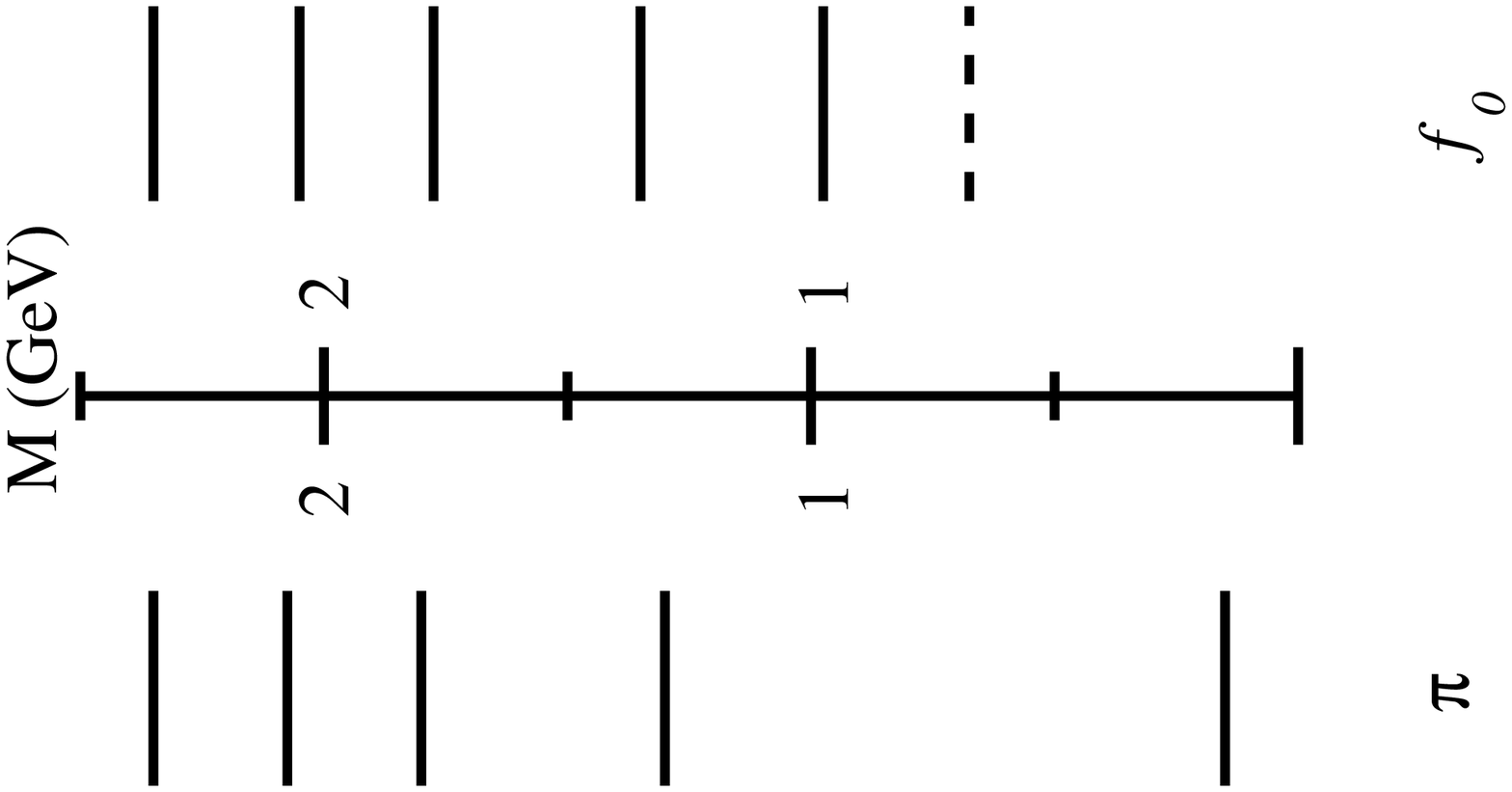}
\end{center}
\caption{Left panel: excitation spectrum of the nucleon (those
resonances which are not yet established are marked by two or one
stars according to the PDG classification). Right panel: pion and
$n \bar n$ $f_0$ spectra. }
\end{figure}

 By definition this effective chiral
 symmetry restoration means the following. In QCD
hadrons with quantum numbers $\alpha$ are created when one applies
an interpolating field (current) $J_\alpha$ with such
quantum numbers on the vacuum $|0\rangle$. Then all
 hadrons that are created by the given interpolator appear
as intermediate states in the two-point correlator,

\begin{equation}
\Pi =\imath \int d^4x ~e^{\imath q x} \langle 0 | T \{ J_\alpha
(x) J^\dagger_\alpha (0) \} |0\rangle, \label{corr}
\end{equation}

\noindent where all possible Lorentz and Dirac indices (specific
for a given interpolating field) have been omitted. Consider two
interpolating fields $J_1(x)$ and $J_2(x)$ which are
connected by a chiral transformation (or by a $U(1)_A$
transformation), $ J_1(x) = UJ_2(x)U^\dagger. $
 Then, if the vacuum
was invariant under the chiral group, $U|0\rangle = |0\rangle,$ it
follows from (\ref{corr}) that the spectra created by the
operators $J_1(x)$ and $J_2(x)$ would be identical. We know that
in QCD one finds $U|0\rangle \neq |0\rangle.$ As a consequence the
spectra of the two operators must be in general different.
However, it happens that the noninvariance of the vacuum becomes
unimportant (irrelevant) high in the spectrum. Then the spectra of
both operators become close at large masses and asymptotically
identical. This  means that chiral symmetry is effectively
restored. We stress that this effective chiral symmetry
restoration does not mean that chiral symmetry breaking in the
vacuum disappears, but that the role of the quark condensates that
break chiral symmetry in the vacuum becomes progressively less
important high in the spectrum. One could say, that the valence
quarks in high-lying hadrons {\it decouple} from the QCD vacuum.

\section{Chiral symmetry restoration and the quark-hadron duality}

 There is
a heuristic argument that supports this idea \cite{CG}. The
argument is based on the well controlled behaviour of the
two-point function (\ref{corr}) at the large space-like momenta
$Q^2 =-q^2$, where the operator product expansion (OPE) is valid
and where all nonperturbative effects can be absorbed into
condensates of different dimensions \cite{SVZ}. The key point is
that all nonperturbative effects of the spontaneous breaking of
chiral symmetry at large $Q^2$ are absorbed into the quark
condensate $\langle \bar q q \rangle$ and other quark condensates
of higher dimension. However, the contributions of these
condensates to the correlation function are proportional to $(1/Q^2)^n$,
where the index $n$ is determined by the quantum numbers of the
current $J$ and by the dimension of the given quark condensate.
Hence, at large enough $Q^2$ the two-point correlator becomes
approximately chirally symmetric. At these high $Q^2$ a matching
with the perturbative QCD (where no SBCS occurs) can be done. 

 Then we can invoke into
analysis a dispersion relation.
Since the large $Q^2$ asymptotics of the correlator
is given by the leading term of
the  perturbation theory,
then the asymptotics of the spectral density, $\rho(s)$ ,at $s \rightarrow \infty$ must
also be given by the same term of the perturbation theory if the
spectral density approaches a constant value (if it oscillates, then
it must oscillate around the perturbation theory value). Hence
both spectral densities $\rho_{J_1}(s)$ and $\rho_{J_2}(s)$ at
$s \rightarrow \infty$ must approach the same value and the spectral
function becomes chirally symmetric. This is definitely true in the
asymptotic (jet) regime where the spectrum is strictly continuous.
The conjecture of ref. \cite{CG} was that may be this is also true
in the regime where the spectrum is still quasidiscrete and saturated
mainly by resonances.

The question arises then what is the functional behaviour that
determines approaching the chiral-invariant regime at large $s$?
One would expect that OPE could help us. This is not
so, however, for two reasons. First of all, we know
phenomenologically only the lowest dimension quark condensate.
But even if we knew all  quark condensates
up to a rather high dimension, it would not help us. This is
because the OPE is only an asymptotic expansion.
While such kind of expansion is very useful in the space-like
region, it does not define any analytical solution which could
be continued to the time-like region at finite $s$.
This means that while the real (correct) spectrum of QCD must
be consistent with OPE, there is an infinite amount of incorrect spectra
that  can also be consistent with OPE. Then, if one wants to get
some information about the spectrum from the OPE side, one needs
to assume something else on the top of OPE. Clearly a success then is
crucially dependent on these additional assumptions, for the recent
activity in this direction  see refs. \cite{OPE,SHIFMAN,GOLTERMAN}.
This implies that in order to really understand chiral symmetry restoration 
 one needs a microscopic insight and theory that would incorporate
{\it at the same time} chiral symmetry breaking and confinement.

\section{Restoration of the classical symmetry in the
semiclassical regime}
A fundamental insight into phenomenon can be obtained from the
semiclassical argument \cite{G5}.
We know that the axial anomaly  as well as
the spontaneous breaking of chiral symmetry in QCD is an effect
of quantum fluctuations of the quark field.
The latter can
generally be seen from the definition of the quark condensate,
which is a closed quark loop.
This closed quark loop explicitly contains a factor $\hbar$. The
chiral symmetry breaking, which is necessaraly a nonperturbative
effect, is actually a (nonlocal) coupling of a quark line with the
closed quark loop, which is a graphical representation of the Schwinger-Dyson (gap)
equation. Hence chiral symmetry breaking in QCD manifestly vanishes in
the classical limit $\hbar \rightarrow 0$.

At large $n$ (radial quantum number) or at large angular
momentum $J$ we know that in quantum systems the {\it semiclassical}
approximation  {\it must} work. Physically this approximation
applies in these cases because the de Broglie wavelength of
particles in the system is small in comparison with the
scale that characterizes the given problem. In such a system
as a hadron the scale is given by the hadron size while the
wavelength of valence quarks is given by their momenta. Once
we go high in the spectrum the size of hadrons increases as well as
 the typical momentum of valence quarks.
This is why a highly excited hadron  can be described semiclassically
in terms of the underlying quark  degrees of freedom.

The physical content of the semiclassical approximation is
most transparently given by the path integral. The contribution
of the given path to the path integral is regulated by the
action $S(\phi(x))$ along the  path $\phi(x)$ (the
fields $\bar \psi,\psi,A$ are collectively denoted as $\phi$) through
the factor $\sim e^{i\frac{S(\phi(x))}{\hbar}}$.
The semiclassical approximation  applies when the
action in the system $S \gg \hbar$.
In this case the whole amplitude (path integral) is dominated by
the classical path $\phi_{cl}(x)$ (stationary point) and those paths that 
are infinitesimally
close to the classical path.  In other words, in the semiclassical case the quantum
fluctuations effects are strongly suppressed and vanish asymptotically.
Then the generating functional can be expanded
in powers of $\hbar$ as

\begin{equation}
W(J) = W_0(J) + \hbar W_1(J) + ...,
\label{f}
\end{equation}

\noindent
where $W_0(J) = S(\phi_{cl}) + J\phi_{cl}$ and $W_1(J)$
represents contributions of the lowest order quantum fluctuations around the
classical solution (determinant of the classical
solution).
The classical path, which is generated by $W_0$,
 is a tree-level contribution
to the path integral
 and keeps chiral symmetries of the classical
Lagrangian. Its contribution is of the order
$(\hbar/S)^0$. The quantum fluctuations contribute at the
orders $(\hbar/S)^1$ (the one loop order, generated by $W_1$),
 $(\hbar/S)^2$ (the two loops order),
etc.
The $U(1)_A$ as well as the spontaneous $SU(2)_L \times SU(2)_R$
breakings  start from the
one-loop order.
However, in a hadron with
large enough $n$ or  $J$, where action is large,  the loop
contributions must be relatively
suppressed and vanish asymptotically.
Then it follows 
that in such systems both
the chiral and $U(1)_A$ symmetries should be approximately restored.
This is precisely what we see
phenomenologically.

While the argument above is solid, theoretically it is
not clear {\it a-priori} whether isolated hadrons still
exist at excitation energies where a semiclassical regime
is achieved.  However, the large $N_c$ limit of QCD,
while  keeping all basic properties of QCD like
asymptotic freedom, confinement and chiral symmetry breaking, allows for a significant
simplification. 
In this limit it is known that all mesons
represent  narrow states.  At the same time the spectrum of mesons is
infinite.  Then one can always excite a meson of any
 arbitrary large energy, which is of any arbitrary large size.
 In such a meson the action $S \gg \hbar$. Hence a description
 of this meson necessarily must  be semiclassical. 
Actually we do not need the exact $N_c = \infty$ limit
for this statement. It can be formulated in the following
way. For any large $S \gg \hbar$ there always exist such
$N_c$ that the isolated meson with such an action does exist
and can be described semiclassically. From the  empirical
fact that we observe  multiplets of chiral
and $U(1)_A$ groups high in the hadron spectrum it follows
that $N_c=3$ is large enough for this purpose. 

\section{A solvable model of the 't Hooft type}

While the argument presented above is general and solid enough,
 a detailed microscopical picture is missing. Then to see
how all this works one needs a solvable field-theoretical model.             
Clearly the model must be chirally
symmetric and contain the key elements, such as confinement and spontaneous
 breaking of chiral symmetry. 
Such a model is known, it is a
generalized Nambu and Jona-Lasinio model (GNJL) with the instantaneous 
Lorentz-vector
confining kernel \cite{Finger:1981gm,Orsay,BR}. This model can be considered as a 
generalization
of the large $N_c$ `t Hooft model (QCD in 1+1 dimensions) \cite{HOOFT} 
to 3+1 dimensions.
In both models the only interaction between quarks is the instantaneous
infinitly raising Lorentz-vector linear potential. Then chiral symmetry breaking
is described by the standard   summation of the valence quarks
self-interaction loops (the Schwinger-Dyson or gap
equations), while mesons are obtained from the Bethe-Salpeter equation for the
quark-antiquark bound states.

An obvious advantage of the GNJL model is that it can be applied in 3+1
dimensions to systems of arbitrary spin.
In  1+1 dimensions there is no spin, the rotational motion of quarks 
is impossible.
 Then it is known that the spectrum represents an alternating
sequence of positive and negative parity states and chiral multiplets never
emerge. In 3+1 dimension, on the contrary, the quarks can rotate and hence
can always be ultrarelativistic
and chiral multiplets should emerge naturally \cite{G2}.
Restoration of chiral symmetry in excited heavy-light mesons has been
previously studied with the quadratic confining potential \cite{KNR}. 
Here we report our  results for excited light-light mesons with
the linear  potential \cite{WG}.

An effective chiral symmetry restoration means that (i) the states
fall into approximate multiplets of $SU(2)_L \times SU(2)_R$ 
and the splittings within the multiplets ( $\Delta M = M_+ -M_-$) 
vanish at $n \rightarrow \infty$ and/or $ J \rightarrow \infty$ ;
(ii) the splitting within the multiplet is much smaller
than between the two subsequent multiplets  \cite{G3,G4,G5}. Note that 
within the present
model the axial anomaly is absent so the mechanism of the
$U(1)_A$ symmetry breaking and restoration is exactly the same as of $SU(2)_L \times SU(2)_R$.

The condition (i) is very restrictive, because the structure of
the chiral multiplets for the $J=0$ and $J>0$ mesons 
is very different \cite{G3,G4}.
For the $J>0$ mesons chiral symmetry requires a {\it doubling} of states
with some quantum numbers.  
Given the  set of  quantum numbers $I, J^{PC}$,
the  multiplets of  $SU(2)_L \times SU(2)_R$ for the $J=0$ mesons are
\begin{eqnarray}
(1/2,1/2)_a  &  :  &   1,0^{-+} \longleftrightarrow 0,0^{++}  \nonumber \\
(1/2,1/2)_b  &  : &   1,0^{++} \longleftrightarrow 0,0^{-+} ,
\end{eqnarray}
%
%
while for the  mesons of even spin, $J>0$,  they are
\begin{eqnarray}
 (0,0)  & :  &   0,J^{--} \longleftrightarrow 0,J^{++}  \nonumber \\
 (1/2,1/2)_a  & : &   1,J^{-+} \longleftrightarrow 0,J^{++}  \nonumber \\
 (1/2,1/2)_b  & : &   1,J^{++} \longleftrightarrow 0,J^{-+}  \nonumber \\
 (0,1) \oplus (1,0)  & :  &   1,J^{++} \longleftrightarrow 1,J^{--} 
\end{eqnarray}
and for odd $J$ they are
\begin{eqnarray}
 (0,0)  & :  &   0,J^{++} \longleftrightarrow 0,J^{--}  \nonumber \\
 (1/2,1/2)_a  & : &   1,J^{+-} \longleftrightarrow 0,J^{--}  \nonumber \\
 (1/2,1/2)_b  & : &   1,J^{--} \longleftrightarrow 0,J^{+-}  \nonumber \\
 (0,1) \oplus (1,0)  & :  &   1,J^{--} \longleftrightarrow 1,J^{++} .
\end{eqnarray}

Restoration of the $U(1)_A$ symmetry would mean a degeneracy of
the opposite spatial parity states with the same
isospin from the distinct $(1/2,1/2)_a $ and $(1/2,1/2)_b$
 multiplets of $SU(2)_L \times SU(2)_R$. Note that within the
 present model there are no vacuum fermion loops. Then since
 the interaction between quarks is flavor-blind the states
 with the same $J^{PC}$ but different isospins from the distinct
 multiplets $(1/2,1/2)_a $ and $(1/2,1/2)_b$ as well as the states
 with the same $J^{PC}$ but different isospins from $(0,0)$ and
 $(0,1) \oplus (1,0)$ representations are  degenerate.

In the Table below  we present  masses (in units $\sqrt{\sigma}$) of
$I=1$ mesons with $J=0,1$ and $J=6$.
A very fast restoration of
both $SU(2)_L \times SU(2)_R$ and $U(1)_A$ symmetries with increasing
$J$ and essentially more slow restoration with increasing of $n$ is observed.
\begin{table}
\begin{tabular}{c@{\hspace*{1.5em}}c@{\hspace*{1.5em}}rrrrrrr}
\hline\hline
\multicolumn{1}{c}{chiral}&\raisebox{-1.5ex}{$J^{PC}$}&
\multicolumn{7}{c}{radial excitation $n$}\\[-1.ex]
\multicolumn{1}{c}{multiplet}&&0\hspace*{0.7em}&1\hspace*{0.7em}&2\hspace*{0.7em}&3\hspace*{0.7em}&
4\hspace*{0.7em}&5\hspace*{0.7em}&6\hspace*{0.7em}\\
\hline
$(1/2,1/2)_a$&$0^{-+}$&0.00&2.93&4.35&5.49&6.46&7.31&8.09\\
$(1/2,1/2)_b$&$0^{++}$&1.49&3.38&4.72&5.80&6.74&7.57&8.33\\
\hline
$(1/2,1/2)_a$&$1^{+-}$&2.68&4.03&5.15&6.14&7.01&7.80&8.53\\
$(1/2,1/2)_b$&$1^{--}$&2.78&4.18&5.32&6.30&7.17&7.96&8.68\\
$(0,1)\oplus(1,0)$&$1^{--}$&1.55&3.28&4.56&5.64&6.57&7.40&8.16\\
$(0,1)\oplus(1,0)$&$1^{++}$&2.20&3.73&4.95&5.98&6.88&7.69&8.43\\
\hline
$(1/2,1/2)_a$&$6^{-+}$&6.88&7.61&8.29&8.94&9.55&10.1&10.7\\
$(1/2,1/2)_b$&$6^{++}$&6.88&7.61&8.29&8.94&9.56&10.1&10.7\\
$(0,1)\oplus(1,0)$&$6^{++}$&6.83&7.57&8.25&8.90&9.51&10.1&10.7\\
$(0,1)\oplus(1,0)$&$6^{--}$&6.83&7.57&8.26&8.90&9.52&10.1&10.7\\
\hline\hline
\end{tabular}
\end{table}

In Fig. 2 the rates of the symmetry restoration against the radial
quantum number $n$ and spin $J$ are shown. It is seen that with the fixed
$J$ the splitting within the multiplets $\Delta M$ decreases asymptotically as
$1/\sqrt n$, dictated by the asymptotic linearity of the radial Regge
trajectories. This property is consistent with the dominance of the free
quark loop logarithm at short distances.
\begin{figure}
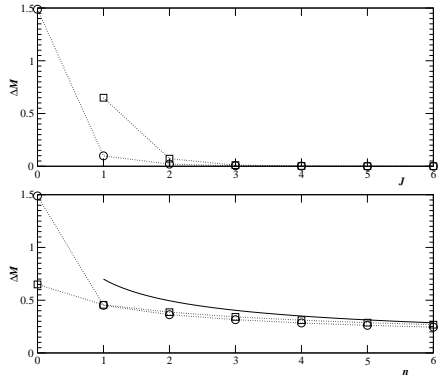

\begin{center}
\includegraphics*[width=0.5\hsize]{delta_m_spin.eps}
\includegraphics*[width=0.5\hsize]{delta_m_radial.eps}
\end{center}
\caption{Mass splittings in units of $\sqrt{\sigma}$ for isovector mesons of the chiral
multiplets $(1/2,1/2)_a$ and $(1/2,1/2)_b$ (circles) and within
the multiplet $(0,1)\oplus(1,0)$ (squares) against $J$ for $n=0$ (top) and
against $n$ for $J=0$ and $J=1$, respectively (bottom).
The full line in the bottom plot is $0.7\sqrt{\sigma/n}$.}
\end{figure}

In Fig. 3 the angular Regge trajectories are shown.
 They exhibit deviations from the linear behavior.
This fact is obviously related to the
chiral symmetry breaking effects for lower mesons.

\begin{figure}
\begin{center}
\includegraphics[width=0.7\hsize]{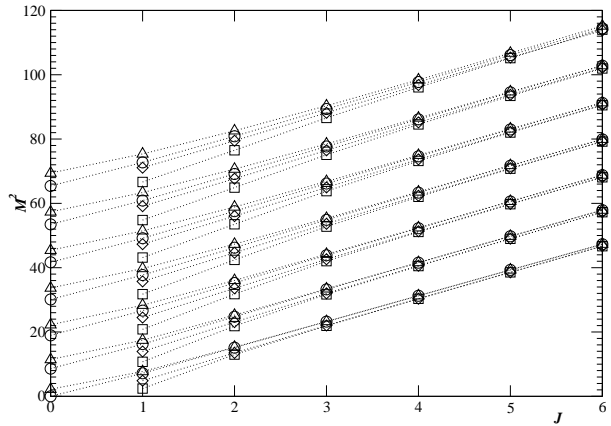}
\end{center}
\caption{Angular  Regge trajectories for isovector mesons
with $M^2$ in units of $\sigma$.
Mesons of the chiral multiplet $(1/2,1/2)_a$ are indicated by circles,
of $(1/2,1/2)_b$ by triangles, and of $(0,1)\oplus(1,0)$
by squares ($J^{++}$ and $J^{--}$ for even and odd $J$, respectively)
and diamonds ($J^{--}$ and $J^{++}$ for even and odd $J$, respectively).
}
\end{figure}

In  the limit $n \rightarrow \infty$ and/or  $J \rightarrow \infty$ 
one observes a complete degeneracy of all multiplets, which means
that the states fall into 
$$ [(0,1/2) \oplus (1/2,0)] \times [(0,1/2) \oplus (1/2,0)]$$
%
representation that combines all possible chiral representations
for the systems of two massless quarks \cite{G4}. This means that in this limit the
loop effects disappear completely and the system becomes classical \cite{G5,G6}.

A few words about physics which is behind these results are in order. In
highly excited hadrons a typical momentum of valence quarks is large.
Consequently, the chiral symmetry violating Lorentz-scalar dynamical
mass of quarks, which is a very fast decreasing function at larger momenta,
becomes small and asymptotically vanishes \cite{G1,G6,G7}. Consequently,
chiral and $U(1)_A$ symmetries get approximately restored. Exactly the
same reason implies a decoupling of these hadrons from the Goldstone bosons \cite{G2,G7}.
Namely, the coupling of the valence quarks to Goldstone bosons is constraint by
the axial current conservation, i.e. it must satisfy the Goldberger-Treiman
relation. Then the coupling constant must be proportional to the Lorentz-scalar
dynamical mass of valence quarks and vanishes at larger momenta. This represents
a microscopical mechanism of decoupling which is required by the general considerations
of chiral symmetry in the Nambu-Goldstone mode \cite{CG2,JPS}.

This work was supported by the Austrian Science Fund (projects P16823-N08 and P19168-N16).



\bibliographystyle{ws-procs9x6}
\bibliography{ws-pro-sample}

\end{document}